# *Deep learning prediction of stress fields in additively manufactured metals with intricate defect networks*

Brendan P. Croom\*, Michael Berkson, Robert K. Mueller, Michael Presley, Steven Storck

Research and Exploratory Development Department, The Johns Hopkins University Applied Physics Laboratory (JHU/APL), Laurel, MD 20723, USA

\* Corresponding author: brendan.croom@jhuapl.edu

*Abstract*

In context of the universal presence of defects in additively manufactured (AM) metals, efficient computational tools are required to rapidly screen AM microstructures for mechanical integrity. To this end, a deep learning approach is used to predict the elastic stress fields in images of defect-containing metal microstructures. A large dataset consisting of the stress response of 100,000 random microstructure images is generated using high-resolution Fast Fourier Transform-based finite element (FFT-FE) calculations, which is then used to train a modified U-Net style convolutional neural network (CNN) model. The trained U-Net model more accurately predicted the stress response compared to alternative CNN architectures, exceeded the accuracy of low-resolution FFT-FE calculations, and was generalizable to microstructures with complex defect geometries. The model was applied to images of real AM microstructures with severe lack of fusion defects, and predicted a strong linear increase of maximum stress as a function of pore fraction. Together, the proposed CNN offers an efficient and accurate way to predict the structural response of defect-containing AM microstructures.

*1.    Introduction*

The ubiquitous presence of porosity defects in additively manufactured metals is known to limit the mechanical properties compared to fully-dense materials during quasistatic and dynamic monotonic mechanical loading [1,2], as well as cyclic fatigue loading [3–5]. During laser powder bed fusion (L-PBF) processing, elaborate networks of voids can form by several mechanisms such as melt pool instabilities, incomplete melting of powder, and entrapped gas porosity [6,7]. Moreover, variation between AM machines, and powder quality contribute to a wide range in observed porosity despite using "standard" build conditions that should result in equivalent, fully dense parts [8]. Consequently, there is an increasing desire to develop tools that can rapidly inspect the defect populations of AM components, and use this information to predict the part's mechanical response.

Currently, the finite element (FE) method is widely used to predict the mechanical performance of AM materials, but these techniques are too computationally expensive to perform high-throughput screening of components. FE models often use as input images of the component's internal porosity obtained by X-ray Computed Tomography (XCT). XCT can resolve internal flaws with a spatial resolutions on the order of 1 to 100 μm [9], which provides the FE model with explicit detail about the defects in ways that cannot be captured with analytical models [1,3,10,11]. This approach is justified by extensive documentation that failure in AM components emanates from large defects that can be detected by XCT, as shown by *in situ* experiments during quasistatic [12,13] and dynamic loading [2]. Similarly, fatigue cracks have been shown to initiate exclusively at large flaws or surface features that can be reliably detected by XCT [14–16]. The capability of this technique was profoundly demonstrated in the recent Sandia Fracture Challenge, where the mechanical response and fracture path was accurately predicted in an AM component with a complex geometry and pore network [17,18]. Similar modeling efforts have shown success in predicting the location of fatigue crack initiation, and are especially promising with regards to capturing interactions between internal porosity and the specimen's free surface [15,19].

Of course, the primary drawback to FE is computational cost, which makes these approaches intractable for high-throughput computations of complex, XCT-derived models. FE models must incorporate highly-refined meshes around pores to ensure convergence, often resulting in models with millions of elements. For example, large-scale FE simulations of numerically-generated tensile coupons with AM-like defects required 790 CPU hours each to solve [10]. A similarly detailed analysis on a fiber-reinforced composite material required 92 hours using 300 cores [20]. Recently, Fast Fourier Transform (FFT)-based FE solvers have been used to evaluate the mechanical response of defect-containing AM metals [21], composites [22] and foams [23]. FFT-based solvers operate directly on images to compute the mechanical response, and exploit the computational efficiency of the FFT to accelerate FE computations. Still, these techniques are inadequate for real-time assessment of XCT microstructures with routine computational equipment.

To further accelerate the stress field computation, we propose the use of convolutional neural network (CNN) machine learning techniques to approximate the FE calculation of stress fields in complex, defect-containing microstructures that are representative of parts produced by AM processes. CNNs are ideally suited for this task, as they have been shown to exhibit exceptional function-approximating capacity that can reproduce the results of FE models, make use of spatially encoded information in images, and can be evaluated substantially faster than conventional or FFT-based FE solvers [24]. For this reason, CNNs have become increasingly popular in evaluating the mechanics of materials with spatially complex microstructures, such as predicting the elastic response of complex composite materials [25], the fracture toughness of digital materials [26], and the strength of complex biological organs [27], among many other applications.

Most relevant to this work, Nie recently designed a CNN called StressNet with an encoder-decoder architecture to predict the stress distribution in cantilever beams with a mean relative stress error of 2% [28]. Their StressNet model applied a series of convolutional filters to images of the beam geometry, and predicted the Von Mises stress for each pixel of the image; these values were compared to the stress on the same model computed by FE. This architecture has recently been extended to predict the mechanical response of fiber-reinforced composites at the microscale [29] and mesoscale [30], and also to predict stress concentrations around microscale pores in a multiscale FE model [31]. Alternative CNN architectures such as generative adversarial network [32] or image colorization networks [33] used to predict stress from microstructural images with varying levels of success. While promising, these studies have exposed several limitations. First, these CNN models have been trained on a manually-designed and/or small datasets with relatively simple geometries, which limited the model's generalizability to AM materials with complex networks of irregularly shaped pores. Second, the stress predictions were relatively poor near closely spaced defects [32], and seemed to struggle with complex, fiber-reinforced composite microstructures [30]; this problem may be related both to the limited CNN depth, as well as limitations in the training dataset. Third, training data were produced using image-based FE techniques of low-resolution microstructure renderings with severely pixelated microstructural features, which would hinder numerical convergence of the FE models [34]. CNNs have also been used to estimate the stress localization in microstructures with relatively low stiffness contrast [25,35], which is an easier problem than modeling the effects of voids on stress localization.

In this paper, we introduce several improvements to the CNN architecture and training methodology, which result in a large improvement in model performance, computational efficiency and generalizability. Namely, these modifications include:

- Training data is generated using a FFT-based FE solver [36], which can produce large training datasets substantially faster than possible with conventional FE solvers. The larger training dataset provides better coverage of possible microstructures, and enhances the generalizability of the CNN prediction. The current training dataset consists of stress fields for 110,000 images of size $384 \times 384\ px$, which is believed to be the largest dataset of its kind. Moreover, the training is performed on randomly generated microstructures, rather than human-designed geometries as in [28].

- As a consequence of the FFT solver's efficiency, we were able to decouple the resolution of the images used for FE analysis from the resolution of the images used to train the CNN model. The high-resolution FE analysis aids with CNN convergence by reducing spurious numerical errors in the training data, and provides the CNN with a degree of "superresolution" to predict the higher-resolution FE results. Prior work has shown that FE-predicted stress fields at discontinuities in low-resolution microstructural images tend to oscillate about the true solution, and thus should be averaged to improve performance [34].

- Finally, we introduce a U-Net style architecture [37] that includes shortcut connections between the encoder and decoder portions of the model. Compared to the StressNet [28] and simple image colorization CNN [33] architectures, this design allows the model to readily incorporate information from multiple length scales into the stress prediction, which enhances the model's performance.

The results of these architecture changes are demonstrated on numerically generated 2D architectures, and impact of these changes are extensively documented through an ablation study. Finally, the trained model is applied to real AM microstructures to elucidate relationships between porosity and mechanical properties.

## 2. Methods

### 2.1. Generation of training data

To train the model, a microstructure generator produced random, 2D porous microstructures that resembled real AM materials (Fig. 1), whose mechanical response were simulated using a FFT-based FE solver. First, a periodic region of interest (ROI) of size ($1.92 \times 1.92\ mm^2$) was defined. Circular pores of random size and location were added to the ROI with radii sampled from an exponential distribution with a scale parameter of 0.095 mm; the location of each pore in the model was uniformly distributed across the ROI. This distribution has previously been shown to describe the pore distribution in 17-4 PH steels produced by AM [10] using XCT data from [8], as well as other materials such as AlSi$_{10}$Mg [38]. No effort was made to prevent overlap between pores. Indeed, it was believed that this would allow further generalization of the CNN model to predict stress in materials with non-circular and/or tightly clustered pore geometries.

The dataset consisted of 110,000 models with random instantiations of 9 to 225 pores, resulting in pore fractions from 0.5% to 30%. These models were exported as binary images with a high resolution of $384 \times 384\ px$ for FE analysis (the "FE resolution"). Subsequently, thin ligaments between pores of size $2\ px$ or smaller were removed from the images; these ligaments were found to impede numerical convergence during FE analysis. As input to the CNN model, the high-resolution binary images were downsampled through local averaging to a resolution of $96 \times$ 96 $px$ (the "CNN resolution"), resulting in small grayscale images. Note that the size of the small grayscale images was carefully selected to be larger than the receptive field of the CNN, as will be discussed further.

The stress fields in the high-resolution images were calculated using a linear-elastic FFT-based FE solver [36] using plane-stress boundary conditions described in Fig. 1. Matrix material was assigned mechanical properties of $E = 200\ GPa$ and $\nu = 0.29$, which were representative of steel. Void material was assigned trivial mechanical properties that were 1/1000[th] those of the matrix; this was to aid with numerical convergence of the high-contrast composite microstructure, and was found to not affect the resulting stress solution. From the FFT solver result, we evaluated the Von Mises stress $\sigma_{vm}$ at every pixel in the $384 \times 384\ px$ image, which was then normalized by $\sigma_{vm}$ of an equivalently loaded homogeneous material (*i.e.*, no voids). These results were locally averaged over $4 \times 4\ px$ regions to create $96 \times 96\ px$ stress maps.

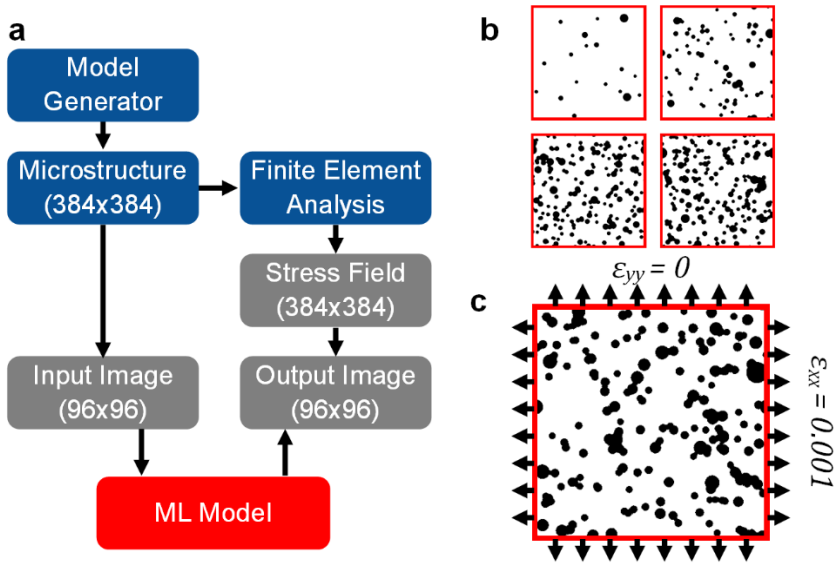

Fig. 1. CNN training framework. (a) Generation of microstructure images and stress fields to train the CNN. (b) Examples of randomly generated microstructures. (c) Boundary conditions used for FEA.

*2.2. CNN design and training*

A modified U-Net style encoder-decoder architecture [37] was developed to predict the stress field in the generated images, and as shown in Fig. 2a. The model accepted padded grayscale images of the porous AM microstructure of size $144 \times 144$ pixels, which were created by tiling the input images according to the periodic boundary conditions used in the FFT solver. The padded images were necessary since the U-Net model employed "valid" convolutions that reduced the image size after each layer. The images were passed through a series of convolutional blocks, which ultimately output an image of the predicted stress field in the central portion of the image, of size $96 \times 96\ px$.

The primary building block of the current CNN architecture is a Squeeze-Excitation Residual Network (SE-ResNet) module as introduced in [39] (Fig. 2b). In contrast to previous work by Sun [30], the SE-ResNet block was used in the encoder, bottleneck and decoder portions of the network. This modification was found to substantially improve the network's predictive capability, and also limit overfitting.

During the encoder portion of the network, an initial convolution of the image was followed by two blocks that contained a SE-ResNet module and a strided convolution to reduce the image resolution. During the bottleneck section of the network, the feature map was passed through a SE-ResNet module and a transposed convolution to increase the image resolution. Next, taking inspiration from U-Net, the decoder portion of the network consisted of a series of blocks in which the input feature map was concatenated with outputs from the corresponding encoder blocks, passed through a SE-ResNet module, and upsampled using a transposed convolution. Lastly, a convolution with a single filter of size $1 \times 1\ px$ consolidated the feature map to create an output of size $96 \times 96\ px$. Except for the last convolution, all convolutions used a filter size of $3 \times 3\ px$, Batch Normalization [40], and ReLU activation [41]. The number of filters in each convolution is defined in Fig. 2a, resulting in a base model with 183,000 parameters. Unless otherwise noted, all assessments of CNN performance refer to this model.

This architecture resulted in a receptive field of size $61 \times 61\ px$, which was used to design the dimensions of the training images. By keeping the receptive field substantially smaller than the size of the padded input images, we prevented the CNN from learning information about the problem's periodicity, and also from extrapolating information from beyond the image boundaries. Indeed, we found that the U-Net made erroneous stress predictions near the image boundary when trained on smaller images, or on large images without padding. Similar errors appear in related work such as [30], where the receptive field of $104 \times 104\ px$ was larger than the image size of $32 \times 32\ px$ and images were not appropriately padded; indeed, nonphysical stress predictions were visible near the image boundary in figures included in [31]. Other related work has shown that inappropriate image padding can worsen predictions of mechanical response [42].

To train the network, the model was implemented in Keras and Python v3.8. The dataset of 110,000 images was split into training, validation and testing sets at a ratio of 90:10:10; the testing data was withheld until training of all CNN models was complete. The loss function was chosen to be pixel-wise mean squared error (MSE), and the model was trained using stochastic gradient descent (SGD) with momentum of 0.9 and batch size of 256 images. The learning rate was ramped up to 0.5 over 50 epochs, and held at 0.5 until the $400^{th}$ epoch; the model training was finished using a learning rate of 0.1 for 50 additional epochs, and 0.01 for 50 more epochs. The model was trained on dual NVIDIA Tesla V100 GPUs, and training completed in roughly 9 hours per model.

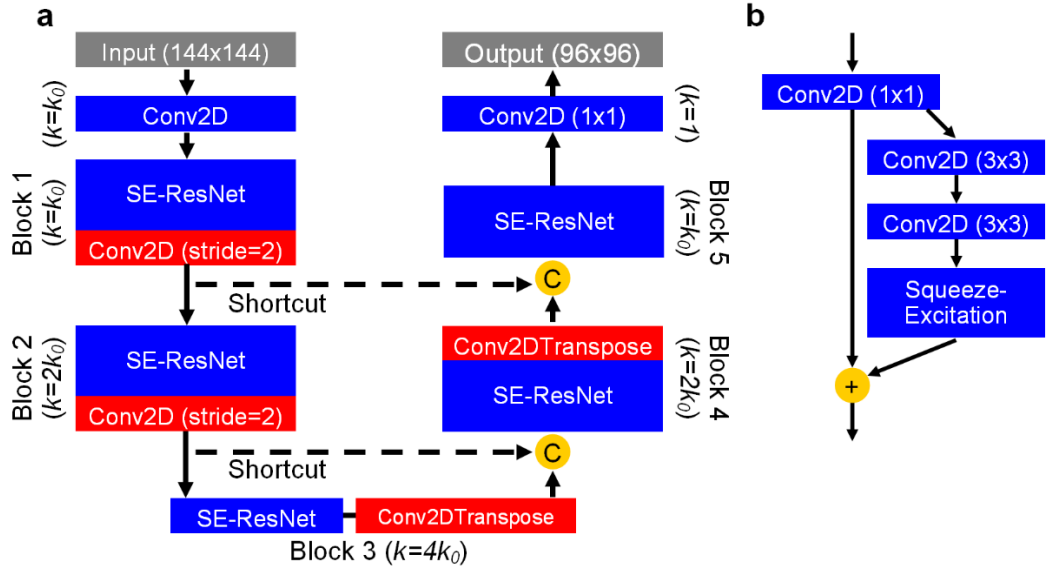

Fig. 2. CNN design. (a) Architecture of proposed modified U-Net. (b) Design of the Squeeze-Excitation Residual Network (SE-ResNet) module. Details of the Squeeze-Excitation module are provided in Ref. [39].

## *2.3. Ablation Study*

To understand the role of key architectural features on the model performance, an ablation study was performed by training similar CNN models with certain features disabled. As appropriate, the number of convolutional filters in each layer was scaled by a uniform factor to maintain similar model sizes of between 180,000 and 190,000 parameters. These variations include (a) removing the shortcut connections (resulting in a model with 182,000 parameters), (b) using a shallower model by eliminating one convolutional layer from the SE-ResNet blocks in Blocks 1-5 (190,000 parameters), (c) training the base model without super-resolution FE data (183,000 parameters), and (d) using zero-padded convolutions that preserve the image size (183,000 parameters).

## *2.4. Comparison to other CNN architectures*

In addition to the U-Net model, we implemented and trained variations of the StressNet [30] and image colorization CNN [33] models. Versions of all three architectures were trained with approximately 46,000, 183,000 and 725,000 parameters, which was achieved by scaling the number of filters in each layer (*i.e.*, changing $k_0$ in Fig. 2). Compared to the U-Net model, the StressNet architecture included fewer convolutional layers in Blocks 1, 2, 4 and 5, and only used SE-ResNet elements in the bottleneck. The image colorization CNN model used a simple architecture of 12 convolutional layers, and did not use any downsampling or upsampling operations to change the filter resolution.

### 2.5. *Application to XCT images of AM microstructures*

To demonstrate the performance of the trained U-Net model on real AM microstructures, we simulated the internal stress distribution of porous Inconel 718 cylinders produced by L-PBF. The cylinders had a diameter of 6 mm, and were manufactured using modified build parameters to induce lack of fusion porosity ranging from 1.5% to 17%. The samples were printed using EOS M290 machine, using powder supplied by EOS. The internal porosity of these samples was imaged using XCT (NorthStar Imaging X-50) at 15 $\mu m$ voxel resolution, and quantified using Volume Graphics software (VGStudioMax 3.4). Central 2D slices of the segmented porosity were exported and cropped to $240 \times 240\ px$ resolution. To compute the stress, we divided the images into overlapping tiles of size $144 \times 144\ px$ that could be evaluated using the trained CNN model.

### 3. **Results**

### 3.1. *Model training and performance*

The training histories of the proposed U-Net as well as the previously developed StressNet [30] and image colorization [33] architectures are presented in Fig. 3 and summarized in Table 1, indicating that the U-Net architecture was able to more accurately reconstruct the FE-measured stress fields. Here and throughout the manuscript, the CNN predictions were normalized by $\sigma_{vm}$ for an equivalently loaded, homogeneous material; as such, the predicted stress fields are dimensionless. After 500 epochs of training, the U-Net achieved a validation MAE of $5.84 \times 10^{-2}$ and MSE of $7.82 \times 10^{-3}$. These values were 1.2 and 2.0 times smaller, respectively, than the StressNet performance with MAE of $9.20 \times 10^{-2}$ and MSE of $2.32 \times 10^{-2}$ for a model of similar size. Additionally, the image colorization model achieved validation MAE of $1.22 \times 10^{-1}$ and MSE of $3.27 \times 10^{-2}$, which were 2.1 and 4.2 times larger than achieved by the U-Net. Additionally, the U-Net and StressNet architectures showed continuous improvement in model performance over 500 epochs, while the image colorization architecture training plateaued after only 100 epochs; this behavior has previously been attributed to the SE-ResNet layers [28,39].

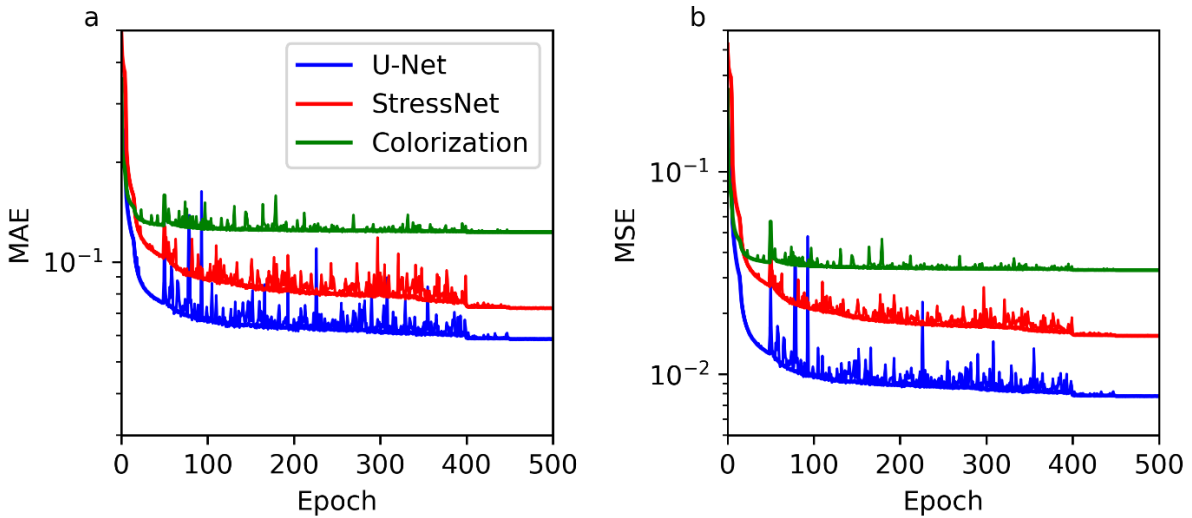

Fig. 3. Training and testing histories for the U-Net, StressNet and image colorization architectures. Results are shown for (a) Mean Absolute Error, and (b) Mean Squared Error. Training data is shown with solid lines, while validation data are shown in dashed lines. Lines are largely coincident with training data, indicating absence of overfitting.

Table 1. Training histories for the baseline U-Net, StressNet and image colorization architectures.

| Model Name | No. Parameters | MSE - Train | MSE - Test | MAE - Train | MAE - Test |
|---|---|---|---|---|---|
| U-Net | 182,677 | $7.78 \times 10^{-3}$ | $7.82 \times 10^{-3}$ | $5.85 \times 10^{-2}$ | $5.84 \times 10^{-2}$ |
| StressNet | 184,765 | $1.55 \times 10^{-2}$ | $1.55 \times 10^{-2}$ | $7.24 \times 10^{-2}$ | $7.22 \times 10^{-2}$ |
| Image Colorization | 170,645 | $3.27 \times 10^{-2}$ | $3.27 \times 10^{-2}$ | $1.23 \times 10^{-1}$ | $1.22 \times 10^{-1}$ |

The performance of the U-Net stress predictions were visually indistinguishable compared to the ground-truth FE results (Fig. 4). Results are presented for six randomly selected microstructures (Fig. 4a). The ground truth and U-Net predicted $\sigma_{vm}$ fields are shown in Fig. 4b and c, and the difference between these results is presented in Fig. 4d with a different color scale to highlight erroneous stress predictions. Since the stress fields were normalized by $\sigma_{vm}$ for an equivalently loaded, defect-free medium, the reported stress values in Fig. 4b-c correspond to the effective stress concentrations due to the defects. The U-Net model accurately reconstructed the very elaborate stress fields that occurred due to irregular void shapes and as well as interactions at multiple length scales between clustered voids. For instance, the model faithfully reproduced stress concentrations in thin ligaments between adjacent voids, and also low stress regions in the shadows of large voids. Notably, similar features were sources of relatively high error in previous work such as [30,32]. Qualitatively, the model performed best in microstructures with fewer voids, *i.e.,* models where the stress had fewer fluctuations.

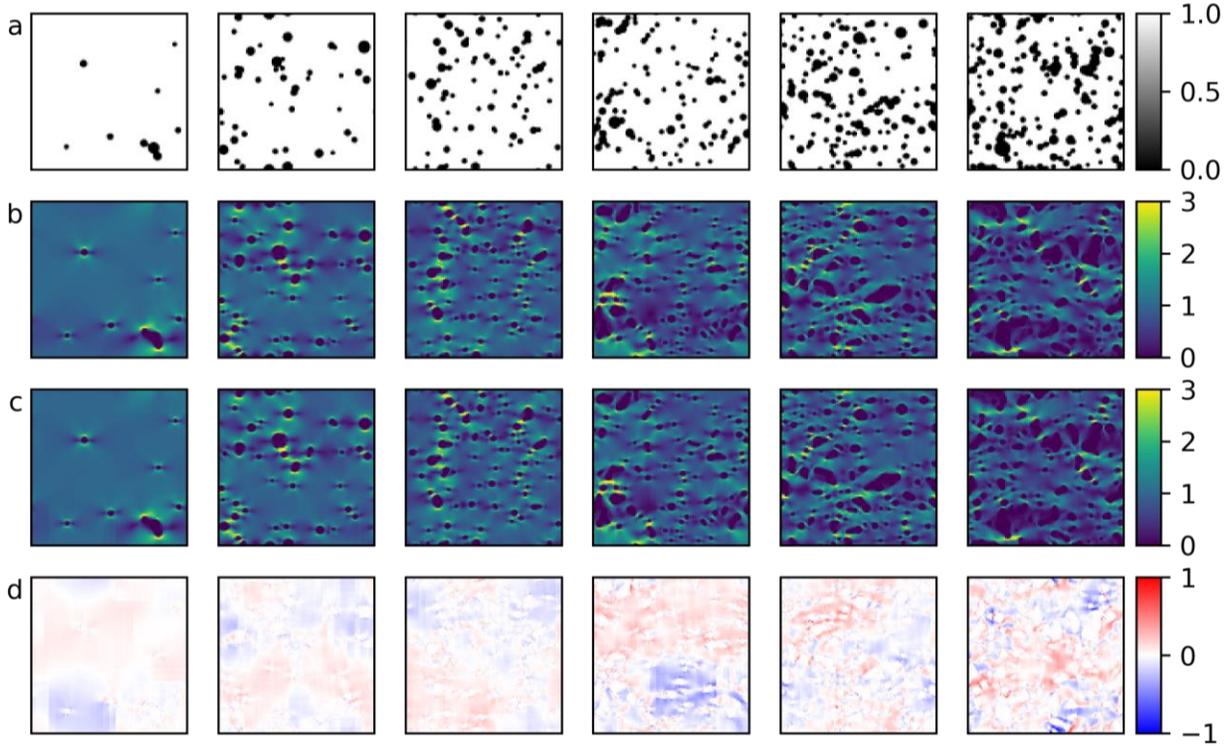

Fig. 4. Evaluation of U-Net model predictions for six different random microstructures. (a) Microstructure image, with pores in black and matrix in white. (b) Ground truth data obtained by FEA calculation. (c) U-Net prediction. (d) Difference between (b) and (c). All images are $96 \times 96\ px$. Loading direction is left to right. Stresses have been normalized by $\sigma_{vm}$ of an equivalently loaded, defect-free medium.

### 3.2. *Ablative study: Role of architecture design choices and FE quality on CNN performance*

To evaluate the role of CNN architecture on the stress predictions, we performed an ablative study where certain features of the CNN were removed from the model. These results are highlighted in Fig. 5a and Table 2, and emphasize the contribution of each CNN architectural feature to the overall model performance.

The largest improvement in performance was attributed to the increased depth of the U-Net model. By removing a convolutional layer from each of the SE-ResNet blocks, the validation MSE increased to $1.81 \times 10^{-2}$. The second most important feature was the use of appropriate image padding and "valid" type convolutions; by removing these features, the validation MSE worsened to $1.66 \times 10^{-2}$. Finally, by removing the shortcut connections that are characteristic of the U-Net architecture, the validation MSE only slightly increased to $8.49 \times 10^{-3}$. By comparison, the StressNet architecture validation MSE was $1.55 \times 10^{-2}$, and the validation MSE for the colorization architecture was $3.27 \times 10^{-2}$. We note that the StressNet architecture was shallower than the proposed U-Net design and did not use appropriate image padding, such that it unsurprisingly achieved similar performance to the *shallow* and *no image padding* U-Net variants. The key architectural difference between the U-Net and the image colorization CNN was the absence of downsampling / upsampling layers, which limited its ability to incorporate information from distant microstructural features into its stress predictions.

The U-Net model substantially outperformed the StressNet and image colorization models with similar or fewer parameters (Fig. 5b and Table 2). By comparing the base U-Net model with 183,000 parameters to equivalently sized StressNet and image colorization models, the U-Net resulted in superior MSE by over a factor of 2 and 4, respectively. Alternately, a U-Net with only 46,000 parameters showed similar validation MSE to a StressNet model with 734,000 parameters. The image colorization model severely underperformed compared to the U-Net model, and showed little improvement as the number of filters in each layer was increased.

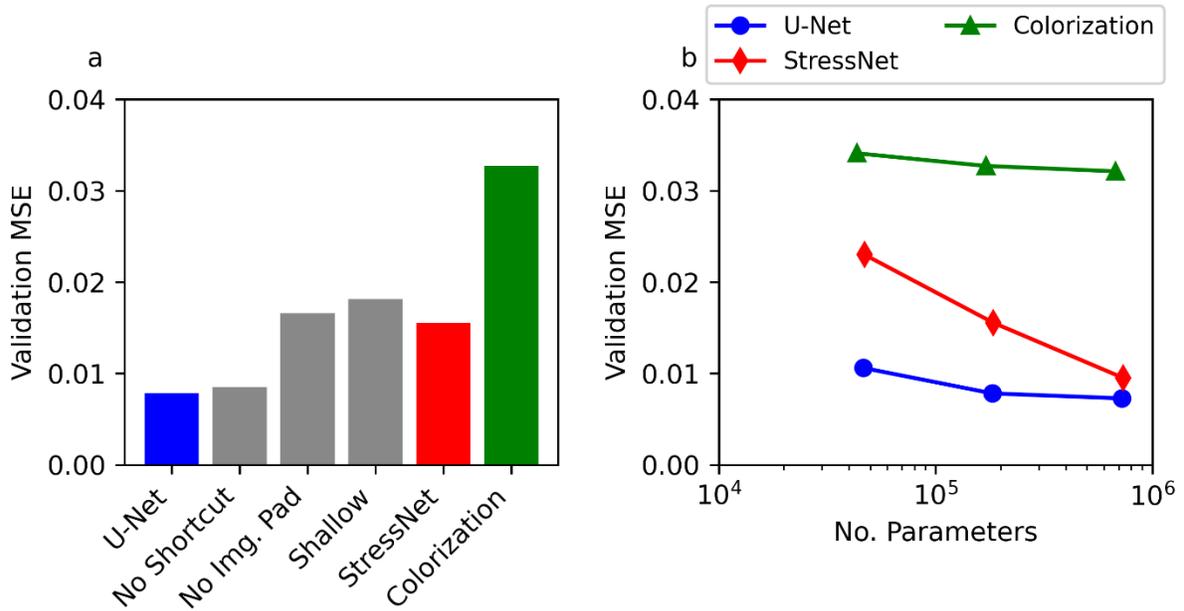

Fig. 5. Role of CNN architecture on model performance. (a) Comparison of U-Net variants and StressNet models with different architectures. (b) Model scaling performance for U-Net and StressNet architectures. All models in (a) have between 170,000 and 190,000 parameters.

Table 2. CNN performance metrics for ablation and model scaling studies.

| Model Name | $k_0$ | No. Parameters | MSE - Train | MSE - Test | MAE - Train | MAE - Test |
|---|---|---|---|---|---|---|
| U-Net - No padding | 16 | 182,677 | $1.64 \times 10^{-2}$ | $1.66 \times 10^{-2}$ | $7.74 \times 10^{-2}$ | $7.76 \times 10^{-2}$ |
| U-Net - No shortcut | 16 | 181,397 | $8.61 \times 10^{-3}$ | $8.49 \times 10^{-3}$ | $6.21 \times 10^{-2}$ | $6.13 \times 10^{-2}$ |
| U-Net - Shallow | 20 | 189,905 | $1.79 \times 10^{-2}$ | $1.81 \times 10^{-2}$ | $8.98 \times 10^{-2}$ | $9.01 \times 10^{-2}$ |
| U-Net | 8 | 46,317 | $1.07 \times 10^{-2}$ | $1.06 \times 10^{-2}$ | $6.88 \times 10^{-2}$ | $6.85 \times 10^{-2}$ |
| U-Net | 16 | 182,677 | $7.78 \times 10^{-3}$ | $7.82 \times 10^{-3}$ | $5.85 \times 10^{-2}$ | $5.84 \times 10^{-2}$ |
| U-Net | 32 | 725,413 | $7.10 \times 10^{-3}$ | $7.27 \times 10^{-3}$ | $5.62 \times 10^{-2}$ | $5.65 \times 10^{-2}$ |
| StressNet | 10 | 46,985 | $2.31 \times 10^{-2}$ | $2.30 \times 10^{-2}$ | $9.25 \times 10^{-2}$ | $9.20 \times 10^{-2}$ |
| StressNet | 20 | 184,765 | $1.55 \times 10^{-2}$ | $1.55 \times 10^{-2}$ | $7.24 \times 10^{-2}$ | $7.22 \times 10^{-2}$ |
| StressNet | 40 | 733,525 | $8.88 \times 10^{-3}$ | $9.52 \times 10^{-3}$ | $5.38 \times 10^{-2}$ | $5.46 \times 10^{-2}$ |
| Image Colorization | 8 | 43,277 | $3.42 \times 10^{-2}$ | $3.41 \times 10^{-2}$ | $1.26 \times 10^{-1}$ | $1.25 \times 10^{-1}$ |
| Image Colorization | 16 | 170,645 | $3.27 \times 10^{-2}$ | $3.27 \times 10^{-2}$ | $1.23 \times 10^{-1}$ | $1.22 \times 10^{-1}$ |
| Image Colorization | 32 | 677,669 | $3.18 \times 10^{-2}$ | $3.21 \times 10^{-2}$ | $1.21 \times 10^{-1}$ | $1.21 \times 10^{-1}$ |

Beyond the CNN architecture, our analysis revealed the importance of high-quality FE training data to ensure accurate predictions. This was evaluated by comparing the ability of each CNN architecture to accurately predict the *maximum stress* in a microstructure (Fig. 6) as a function of the FE resolution compared to the CNN resolution. Notably, the maximum stress identified the ability of the CNN to capture localized stress, which is a much more challenging prediction than the global nature of the MSE metric. When trained on high-resolution FE data, the U-Net, StressNet and image colorization architectures predicted the maximum stress with an accuracy of $R^2 = 0.90$, 0.81 and 0.71, respectively (Fig. 6a-c). In contrast, when trained on low-resolution FE data, the U-Net, StressNet and image colorization architectures achieved $R^2 = 0.44$, 0.43 and 0.36, respectively. Thus, while the proposed U-Net architecture outperformed the other architectures in all cases, the accuracy of all models was strongly limited by the quality of the FE data. In particular, the low-resolution FE data exhibited relatively large numerical errors, and thus could not be used to train models that accurately predicted the maximum stress.

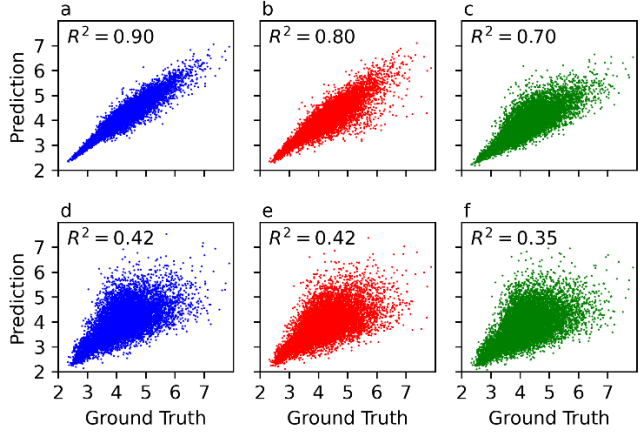

Fig. 6. Accuracy of CNN model predictions of the *maximum stress* in each image. CNN predictions are compared to ground truth values for (a,d) U-Net, (b,e) StressNet and (c,f) image colorization architectures. Models are trained with (a-c) FE data with 4 × 4 superresolution compared to the microstructure images, and (d-f) FE data at the same resolution as the images. The "ground truth" is established using FE data with 4 × 4 superresolution compared to the CNN resolution.

### 3.3. *Application to XCT images of real AM microstructures*

After training the U-Net model and verifying its performance on synthetic microstructures, the model was used to predict the stress distribution in real AM microstructures. Predicted stress fields are shown for six representative lack of fusion microstructures with pore fractions up to 9.51% (Fig. 7). As porosity increased, the stress fields showed more intense variation, as well as increased spatial complexity. For instance, interactions between voids were limited in Fig. 7a, which was due to the relatively small size and large distance between voids. However, as the size and quantity of voids increased, we observed stronger interactions between voids. Higher stresses were observed at thin ligaments of material between adjacent voids, and were amplified near sharp, notch-like voids. Additionally, we observed shadowing effects where large, flat voids reduced the stress in adjacent material in the loading direction. In fact, the dimensions of these shadowing effects were very large in Fig. 7f, and even appeared to be truncated at distances similar to the CNN's receptive field size.

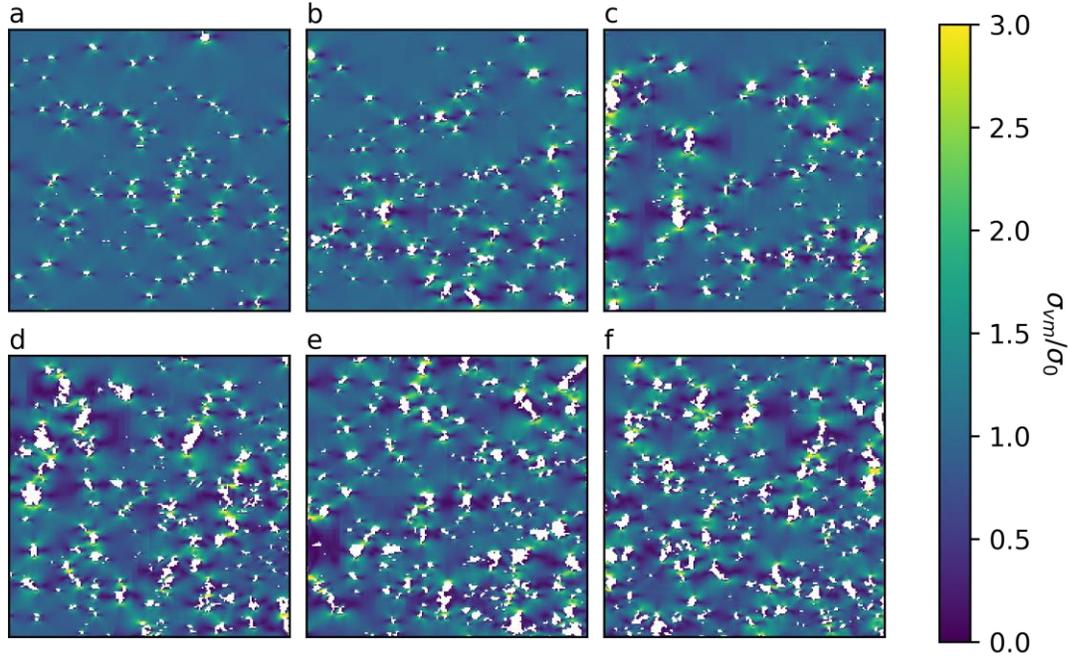

Fig. 7. U-Net predicted stress fields in real AM microstructures with lack of fusion defects. Results are shown for microstructures with pore fractions of (a) 1.85%, (b) 3.57%, (c) 4.60%, (d) 7.94%, (e) 8.82%, and (f) 9.51%. All images are $192 \times 192\ px$, with pores shown in white. Loading and build directions are horizontal.

In aggregate, the CNN-predicted stress fields showed that the *maximum* stress within the field of view increased approximately linearly with the pore fraction (Fig. 8a). In the sample with the lowest porosity of 0.65%, the CNN predicted a normalized maximum stress of 3.61, which increased to 6.15 for a sample with 11.2% porosity. However, there was substantial variation between samples with similar porosity values, as indicated by the low $R^2 = 0.45$. Indeed, the second highest predicted stress was observed in a sample with only 2.49% porosity. As such, this emphasized the roles of pore shape, size and local interactions on predicting the mechanical response of AM materials in ways that could not be explained by global measures of porosity.

In order to evaluate the accuracy of the CNN predictions, the mechanical response of the microstructures was also calculated using the FFT-FE solver. The 90th, 99th and 100th percentiles of the stress fields were compared at different percentiles for each of the 16 microstructure images (Fig. 8b). In general, these results showed that the CNN accurately predicted the stress up to the 99th percentile, but appeared to underestimate the maximum stress. We note that these calculations were performed at the same resolution as the XCT image (*i.e.,* no superresolution), and therefore may exhibit relatively large numerical errors.

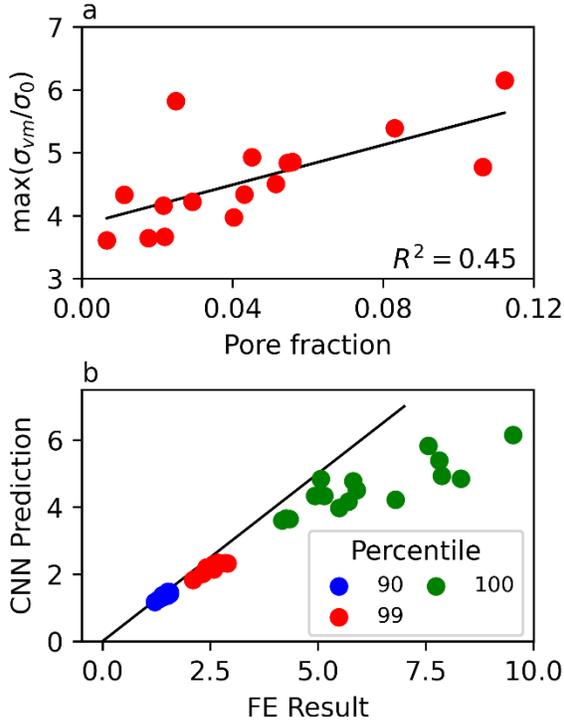

Fig. 8. Summary statistics of CNN predicted stress fields on real AM microstructures. (a) CNN-predicted maximum normalized stress as function of pore fraction. (b) Comparison of CNN predicted and FFT-FE calculated normalized stress at 90th, 99th and 100th stress percentiles.

## 4. *Discussion*

### 4.1. *Comparison of stress fields predicted by CNNs and FE analysis*

The modified U-Net CNN architecture outperformed the previously developed StressNet and image colorization CNNs in ability to predict the stress fields around pore-like defects. An ablative study revealed that this benefit primarily originated from increased CNN depth, better treatment of the model's boundary conditions, and the incorporation of image data from multiple length scales. Together, these changes allowed the U-Net to more accurately replicate the FE stress measurements with fewer parameters than other models.

Compared to conventional and FFT-based FE, the proposed CNN framework offered substantial improvements in speed and accuracy (Table 3). On images of size $96 \times 96\ px$, the CNN evaluated faster than the conventional FFT-FE model (which was already substantially faster than an equivalent conventional FE model) by a factor of 40. This performance benefit improved with even larger images: for example, the stress fields in Fig. 7 were each evaluated in 0.1 seconds, compared to 50 seconds for the FFT-based analysis; this comparison was performed on an 8-core laptop without GPU acceleration, which would have further improved the CNN performance.

More significantly, we also demonstrated that the stress fields predicted by the CNN model could also be *more accurate* than those predicted by FE techniques at equivalent resolution. This benefit primarily originated from training on "super-resolution" FE models with refined meshes, which mitigated numerical convergence errors in the training dataset [34]. Thus, the CNN could more closely approximate the true stress signal in the FE results, and was exposed to fewer artifacts associated with coarse meshes. Given poor training data, the CNN would otherwise learn to mimic both the underlying stress field signal *and* the numerical error in FE model. Indeed, this explained the poor predictive capability of the U-Net and StressNet models when trained on low-resolution FE data (Fig. 6d-f).

Table 3. Performance of CNN model against FFT-FE model. Speed is evaluated as time to compute stress field in a single image of size $96N \times 96N$. Mean Squared Error (MSE) is evaluated on pixel-wise basis against a model computed with $N = 16$.

| Analysis type | Image Resolution factor, N | Evaluation time (s) | MSE against $N = 16$ superresolution FFT-FE |
|---|---|---|---|
| FFT-FE | $1 \times 1$ | 0.8 | $1.64 \times 10^{-1}\ MPa$ |
| FFT-FE | $2 \times 2$ | 5.4 | $2.84 \times 10^{-2}\ MPa$ |
| FFT-FE | $4 \times 4$ | 39 | $4.50 \times 10^{-3}\ MPa$ |
| U-Net CNN | $1 \times 1$ | 0.02 | $9.76 \times 10^{-2} MPa$ |

*4.2.   Application to high-throughput inspection of AM materials*

It is widely established that AM metals are sensitive to void-like defects during quasistatic and fatigue loading. While XCT has been extensively used to identify and measure defects formed during the AM process, there has been no feasible technique to *efficiently* simulate the stress distribution around defects during mechanical loading. Compared to conventional FE techniques, the proposed U-Net CNN model offers a very efficient way to calculate the local stress field around large defects. Importantly, the U-Net model captures interactions between voids at multiple length scales, and provides high-fidelity predictions of maximum stress in the imaged material. Additionally, the model appears to provide reasonable predictions of stress around circular and noncircular defects that are commonly encountered in AM. Once trained, it is possible to evaluate the stress response on large microstructural images using ordinary computational equipment in real time.

We envision using the trained U-Net model to quantitatively inspect defect-containing AM components that have been imaged non-destructively by XCT. Given the speed at which the CNN model can be evaluated compared to acquire and process XCT data (less than 1 second to evaluate the CNN, versus 10-100 minutes to acquire image using laboratory XCT equipment), this analysis can be performed at minimal cost, while providing valuable predictions of the mechanical response.

Moreover, the U-Net approach can be applied to images of defects obtained by *any* imaging modality beyond XCT. For example, the AM community has prioritized the development of in process monitoring techniques that can identify the formation of defects *during the build process*, which would eliminate or minimize the use of costly XCT post-build inspection. For instance, several researchers have demonstrated the ability to identify the formation of defects using infrared cameras [43,44], optical cameras [45–48], and hyperspectral sensors and/or photodiodes [49].

Output from these monitoring techniques can be converted to images of the porosity, which can then be interpreted using the U-Net model.

For application to real AM microstructures, the proposed U-Net encounters two limitations. First, real AM defects exhibit complex 3D geometries, so the U-Net must be modified to process 3D images. As such, the primary bottleneck will be the generation of adequate 3D, high-resolution training data. To this end, the efficient FFT-FE solver offers a feasible pathway to generate training data. Additionally, the CNN model and/or FE image resolution may be reduced using sensitivity studies that consider tradeoffs between resolution and FE/CNN accuracy.

The second limitation is the generalizability of the CNN to real AM defect geometries, particularly since it was trained on synthetic microstructures composed of uniformly distributed circular defects. As demonstrated in Fig. 7, actual lack of fusion defects exhibited highly irregular shapes, and were spatially clustered. To address this deficiency, there is a need to accurately characterize and generate high-fidelity AM microstructures that can be used to train the U-Net. Alternatively, representative microstructures could be extracted from high-resolution XCT scans, and possibly augmented with synthetic data.

## 5. *Conclusion*

A modified U-Net style CNN model has been designed and trained to predict the stress response of porous, defect containing metal microstructures. This architecture showed superior performance in reproducing stress fields calculated by FE compared to previously proposed CNN models. The model predicted the *maximum stress* in porous solids with a correlation coefficient of $R^2 = 0.90$, compared to a value of 0.81 and 0.71 for simpler architectures, and also captured complex patterns in the stress field due pore interactions. Key conclusions of this work include:

- The CNN was successfully trained using data generated with an efficient FFT-based FE solver, which allowed for the development of a large training database of stress fields calculated on high-resolution images. We trained the U-Net model on FE data calculated on 100,000 images with $4 \times 4$ higher resolution than the CNN training images, allowing the CNN model to achieve super-resolution performance. In fact, the accuracy of the CNN model trained on the $4 \times 4$ FE data exceeded the accuracy of raw FE data calculated from images at their native resolution.
- Beyond the important benefit of super-resolution training data, an ablative study of the CNN architecture also indicated that a combination of CNN depth, shortcut connections, and appropriate boundary conditions enhanced the model's performance. Together, these changes allowed the model to capture interactions between defects at multiple length scales, such as stress concentrations around pores and interactions between pores. Notably, these features were poorly predicted by previous CNNs such as StressNet [28] and image colorization [33] models.
- Lastly, the performance of the U-Net was demonstrating by calculating the stress field in defect-containing, porous AM structures. We applied the CNN to XCT-acquired images of porosity in parts with severe lack of fusion defects, and demonstrated that the maximum stress in the part increased linearly with the pore fraction. The stress fields of large images were evaluated in less than 1 second using ordinary computing equipment, which was multiple orders of magnitude faster than the FFT-FE solver. The predicted stress fields strongly correlated with the FFT-FE measured stresses, confirming the model's generalizability to real AM microstructures with complex defect geometries.

*CRediT authorship contribution statement*

**Brendan P. Croom:** Conceptualization, Methodology, Investigation, Visualization, Writing, **Michael Berkson**: Investigation, **Robert K. Mueller**: Investigation, **Michael Presley**: Investigation, **Steven Storck**: Investigation.

*Acknowledgements*

This work was supported by the JHU/APL Independent Research and Development Program.